\documentclass[12pt]{article}
\usepackage{graphicx}
\usepackage{epsfig}
\begin{document}
\baselineskip=5.5mm
\newcommand{\be} {\begin{equation}}
\newcommand{\ee} {\end{equation}}
\newcommand{\Be} {\begin{eqnarray}}
\newcommand{\Ee} {\end{eqnarray}}
\renewcommand{\thefootnote}{\fnsymbol{footnote}}
\def\a{\alpha}
\def\b{\beta}
\def\g{\gamma}
\def\G{\Gamma}
\def\d{\delta}
\def\D{\Delta}
\def\e{\epsilon}
\def\k{\kappa}
\def\l{\lambda}
\def\L{\Lambda}
\def\t{\tau}
\def\om{\omega}
\def\Om{\Omega}
\def\s{\sigma}
\def\lg{\langle}
\def\rg{\rangle}
\noindent
\begin{center}
{\Large
{\bf
Aging in a free-energy landscape model for glassy relaxation
}}\\
\vspace{0.5cm}
\noindent
{\bf Gregor Diezemann} \\
{\it
Institut f\"ur Physikalische Chemie, Universit\"at Mainz,
Welderweg 11, 55099 Mainz, FRG
\\}
\end{center}
\vspace{1cm}
\noindent
{\it
The aging properties of a simple free-energy landscape model for the primary
relaxation in supercooled liquids are investigated.
The intermediate scattering function and the rotational correlation functions are
calculated for the generic situation of a quench from a high temperature to below the
glass transition temperature.
It is found that the re-equilibration of molecular orientations takes longer than for
translational degrees of freedom.
The time scale for re-equilibration is determined by that of the primary relaxation
as an intrinsic property of the model.
}

\vspace{0.5cm}

\vspace{1cm}

\section*{I. Introduction}
If glassy materials are quenched from above their glass transition temperature deep
into the glassy state, they usually do not reach equilibrium on experimentally
accessible time scales.
The dynamics in this out-of-equilibrium situation, the so-called aging phenomena,
have been investigated for a long time, for reviews see e.g.
refs.\cite{Struik, Hodge94}.
If the system reaches equilibrium, i.e. if the quench is not performed at a too low
temperature, the aging behavior often can be well described in terms of the
Tool-Narayanaswami-Moynihan model\cite{TNM}.
This model takes into account the non-linearity and the time-dependence of the
relaxation time via the introduction of a reduced time and a fictive temperature.

In the recent past, many theoretical investigations of the aging behavior of glassy
systems have been undertaken and some rigorous results regarding both, the detailed
time-dependences of two-time correlation functions (CFs) and the violations of the
fluctuation-dissipation theorem (FDT) have been derived for several 
models\cite{CR03}. 
These violations can in favourable cases be used for the definition of a 
so-called effective temperature\cite{CKP97}. 
Also some computer simulations on model supercooled liquids have been performed and 
an effective temperature could be defined\cite{kobbarrat, barrat03}. 
Apart from theoretical considerations a number of experimental investigations of the
violations of the FDT have been performed in the recent past, in particular on 
glassforming liquids\cite{israeloff99}, polymers\cite{BCG03}, colloidal 
glasses\cite{AG04} as well as spin glasses\cite{HO02}. 
In all these systems, strong violations of the FDT have been observed and in some
cases an effective temperature as a function of the time elapsed after the initial
quench into the glassy phase has been determined. 
For glass forming liquids, the effective temperature relaxes to the bath temperature 
for not too deep quenches\cite{israeloff99, BCG03}. 

Without referring to an effective temperature explicitly, it has been found
experimentally that the aging behavior of volume and enthalpy in a glassy polymer may
not be the same\cite{SSP01}. 
A detailed study of the aging properties of polystyrene has been performed by
Thurau and Ediger\cite{mark02}, who clearly showed that probe rotation and probe
translation show different behaviors in this example. 
It was found that after a temperature jump into the glassy phase, the relaxation
towards equilibrium of probe rotation takes longer than the corresponding
re-equilibration of probe translation, if the final temperature of the quench is low
enough. 
In particular, the temperature dependence of the re-equilibration times for rotations
was found to be stronger than that of translational motions. 
The results could consistently be interpreted in terms of spatially heterogeneous
dynamics, which is now well established to be at the origin of many of the peculiar
features of the relaxation in glass forming liquids, for reviews
see refs.\cite{hetero,mark.het}. 
Thurau and Ediger did not attempt to quantitatively discuss the relation between the 
re-equilibration times and the time scale of the primary relaxation at the final
temperature of the quench. 
Such an analysis has been performed in detail for a number of glass forming liquids 
by dielectric spectroscopy\cite{lunki05} and it was shown unequivocally that the 
aging behavior in these examples is determined by the structural $\a$-relaxation. 

In the present paper, I will discuss the aging properties of glass forming liquids
within a simple free-energy model that has been used previously to describe the
heterogeneous relaxation in equilibrium\cite{elm97, elm98, elmb99}. 
The idea of this model is simple. 
It is assumed that the primary relaxation in a supercooled liquid is thermally
activated and proceeds via transitions among a large number of metastable states,
characterized by their free-energy. 
This idea is in accord with various proposed scenarios of glassy relaxation, such as
the one of a random first order transition\cite{XW01}. 
In order to describe the rotational or translational motion of tagged particles, it
is assumed that each transition is accomponied by a particle rearrangement, giving
rise to changes in both, the orientation and the position of the particle considered. 
These assumptions along with a mean jump distance and a mean rotation angle are
enough to compute relevant CFs, such as the intermediate scattering function or
rotational CFs. 
Due to the different ways various CFs average over the distribution of free-energies
a number of hitherto unresolved puzzling features of the $\a$-relaxation could be
explained in a simple way\cite{elm98}. 
This includes the breakdown of the Stokes-Einstein relation and the differences in 
the stretching observed in different experiments probing the reorientational or 
translational dynamics.  
Here, I will concentrate on a qualitative description of aging experiments and show
that the results from model calculations capture the most prominent features observed 
experimentally. 

The paper is organized as follows.  
In the next section, the free-energy model is briefly reviewed and it is described
how to calculate CFs in an aging system. 
In Section III the aging behavior of the intermediate scattering function and of the
rotational CFs is discussed. 
For comparison, in Section IV the aging-behavior of rotational CFs is discussed in 
the framework of models for environmental fluctuations\cite{AU67} or exchange 
models\cite{sill96}. 
Finally, the conclusions are presented in Section V. 

\section*{II. Aging behavior in the free-energy landscape model}
The aging behavior of supercooled liquids depends on the final temperature in a 
temperature jump experiment. 
This is because if the quench is performed to a temperature not too far below the 
glass transition temperature, the system can reach equilibrium on an accessible time 
scale since the glass transition temperature is merely a convenient way of 
classifying the time scale of the primary relaxation\cite{EAN96}. 
For deep quenches the re-equilibration usually cannot be monitored experimentally 
because the relaxation time is extremely long. 
This is one reason why sometimes models exhibiting a true transition into a glassy 
phase are considered when discussing the aging behavior.
As will become clear in the present Section, in the free-energy landscape model the 
system always reaches equilibrium for long times and the re-equilibration time scale 
is determined by the $\a$-relaxation time at the final temperature of the quench. 

In the following, I will utilize the free-energy landscape model for the caculation 
of translational and the rotational two-time CFs\cite{elm98} after a temperature 
jump. 
In case of the reorientational motion, the CF is obtained by correlating the 
orientation-dependent interactions relevant in the experiment considered at two 
times.  
The transformation properties of these interactions determine the so-called 
rotational CF $g_L(t,t_w)$, 
\be\label{gl.t}
g_L(t,t_w)=\lg P_L(\cos{(\theta(t))}P_L(\cos{(\theta(t_w))}\rg 
\ee
Here, $P_L(x)$ denotes the Legendre polynomial of rank $L$. 
For instance, in dielectric spectroscopy $g_1(t,t_w)$ and in NMR or depolarized 
light scattering $g_2(t,t_w)$ are measured. 
Here, $t_w$ denotes the waiting time, i.e. the time that has elapsed after the 
initial preparation of the system before the beginning of the experiment. 
In all later calculations, the initial preparation corresponds to a quench from a 
high temperature to the working temperature. 

If one is interested in translational motions of tagged particles, one naturally 
considers the incoherent intermediate scattering function\cite{squires}:
\be\label{Sq.t}
S_q(t,t_w)=\lg e^{i{\bf q}{\bf r}(t)}e^{-i{\bf q}{\bf r}(t_w)}\rg
\ee
where ${\bf r}(t)$ denotes the position of the particle and ${\bf q}$ is the 
wave-vector. 
Additionally, I have already assumed that the system is isotropic, $q\!=\!|{\bf q}|$. 

In the free-energy model, a free-energy $\e$ is associated with each of the 
exponentially large number of basins or metastable states of the 
system\cite{CR03,SW83}. 
The $\a$-relaxation is modeled using a master equation\cite{vkamp81} for the 
conditional probability to find the system in a metastable state characterized by 
$\e$ at time $t$, given that one had $\e_0$ at $t_0$: 
\be\label{ME.eps}
{\dot G}(\e,t|\e_0,t_0)= -\int\!d\e'\k(\e'|\e)G(\e,t|\e_0,t_0)
                         +\int\!d\e'\k(\e|\e')G(\e',t|\e_0,t_0) 
\ee
Here, $\k(\e'|\e)$ denotes the rate for a $\e\to\e'$-transition, which is assumed to 
be time-independent for simplicity. 
Consequently, $G(\e,t|\e_0,t_0)$ depends only on the difference of the times 
involved, $G(\e,t|\e_0,t_0)\!=\!G(\e,t-t_0|\e_0)$. 
In ref.\cite{elm98}, we considered different choices for the transition rates 
$\k(\e'|\e)$ with only quantitative differences in the results. 
Therefore, in the present calculations, I will solely consider the following simple 
choice for the transition rates, denoted as globally connected model in 
ref.\cite{elm98}.  
It is assumed that every state can be reached by a transition and that the
destination state is chosen at random according to a prescribed density of states, 
denoted by $\eta(\e)$. 
The transition out of a given state is assumed to be thermally activated with a rate 
given by $\k(\e)\!=\!\k_\infty e^{\b\e}$, where a common activation energy has been 
absorbed in the prefactor and $\b\!=\!(k_BT)^{-1}$. 
This means, all calculations will be restricted to the simple case of 
transition rates
\be\label{kap.eps}
\k(\e'|\e)=\k_\infty\eta(\e')e^{\b\e}
\ee
Furthermore, the density of states will be chosen to be Gaussian with zero mean and 
variance $\s$ throughout, $\eta(\e)\!=\!{1\over\sqrt{2\pi}\s}e^{-\e/(2\s^2)}$. 
The same transition rates have also been used by Dyre in his 'energy master 
equation'\cite{jeppe}. With this choice, the system reaches equilibrium for long 
times and the equilibrium probabilities are given by:
\be\label{p.eq.eps}
p^{eq}(\e)=\lim_{t\to\infty}G(\e,t|\e_0,t_0)
={1\over\sqrt{2\pi}\s}e^{-(\e-{\bar\e})/(2\s^2)}
\quad\mbox{with}\quad {\bar\e}=-\b\s^2
\ee
The distribution of equilibrium probabilities thus has a 
temperature-independent width and a mean that scales like the inverse temperature. 
In the inherent structure approach to the classification of the potential 
energy landscapes of supercooled liquids\cite{SW83} it is usually found that the 
inherent structure energies are distributed according to a Gaussian\cite{DH03}. 
However, the transitions among different inherent structures are more complicated 
than assumed in eq.(\ref{kap.eps}), see e.g. ref.\cite{APRV98}.  

The main idea of the present model consists in the coupling of the reorientational 
and translational motion of a tagged particle to the $\a$-relaxation as follows.
The transitions among the metastable states are the only possible way for particle 
rearrangements (apart from vibrational motion within a state). 
Therefore, it is assumed that the particle considered changes its position as well as 
its orientation by a certain amount whenever a transition takes place. 
In order to model the dynamics of either translations or rotations, the composite 
Markov process $\{\e(t),a(t)\}$ with $a(t)$ denoting the orientation $\Om(t)$ or the 
position ${\bf r}(t)$ is considered\cite{elm97,elm98}. 
To this end, one has to deal with the master equation for the combined conditional 
probabilities $G(\e,a,t|\e_0,a_0,t_0)$ and the transition rates $W(\e',a'|\e,a)$.  
For the latter, the simplest possible choice will be considered which means that 
only reorientations with a mean angle $\d\Om$ and only translations with a mean jump 
distance $\d R$ are allowed. 

The solution of the corresponding master equation is outlined in Appendix A for 
convenience of the reader, for more details see ref.\cite{elm98}.
Here, it suffices to note that it is assumed that initially there are no correlations
between the populations $p(\e,0)$ and $p(a,0)$ and the latter are given by the 
equilibrium populations $p^{eq}(a)$, cf. eq.(\ref{p0.eps.a}).
The two-time CFs defined in eqns.(\ref{gl.t}) and (\ref{Sq.t}) are given by:
\Be\label{C.Q.L}
g_L(t,t_w)=&&\hspace{-0.3cm}\int\!d\e\!\int\!d\e'G_L(\e,t|\e')p(\e',t_w)
\nonumber\\
S_Q(t,t_w)=&&\hspace{-0.3cm}\int\!d\e\!\int\!d\e'G_Q(\e,t|\e')p(\e',t_w)
\Ee
Here, the Green's functions $G_L(\e,t|\e')$ and $G_Q(\e,t|\e')$ are solutions of 
eq.(\ref{Gl.Gq.ME}) and $Q\!=\!(q\!\cdot\!\d R)$. 
For long waiting times, it is evident from eq.(\ref{p.eq.eps}) that
\be\label{p.t.to.infty}
p(\e,t_w)\to p^{eq}(\e)\quad{\rm for}\quad t_w\to\infty
\ee
because of $p(\e,t_w)\!=\!\int\!d\e'G(\e,t_w|\e')p(\e',0)$ and 
$\int\!d\e'p(\e',0)\!=\!1$. 
The CFs then reduce to those in equilibrium, 
\be\label{C.Q.L.eq}
g_L^{eq}(\t)\!=\!\int\!d\e\!\int\!d\e'G_L(\e,\t|\e')p^{eq}(\e')
\quad{\rm and}\quad
S_Q^{eq}(\t)\!=\!\int\!d\e\!\int\!d\e'G_Q(\e,\t|\e')p^{eq}(\e')
\ee
the former of which has been discussed in detail in ref.\cite{elm98}.

In order to describe an aging experiment, one has to fix the initial populations.
Experimentally, one usually performs a temperature jump from a 'high' temperature 
above the glass transition temperature $T_g$ to below $T_g$.
In the present paper, I will consider only quenches from $\b\!=\!0$ ($T\!=\!\infty$) 
to the working temperature, which means: 
\be\label{p.0} 
p(\e,0)=\eta(\e)={1\over\sqrt{2\pi}\s}e^{-\e/(2\s^2)}
\ee 
Due to the fact that equilibrium is reached in the long run, cf. 
eq.(\ref{p.t.to.infty}), all aging effects are of a transient nature.
Note that $\b\!=\!0$ in the present context only means that the transition rates in 
eq.(\ref{kap.eps}) are no longer thermally activated, i.e. one has 
$\k_{(\b=0)}(\e'|\e)=\k_\infty\eta(\e')$ independent of the initial state. 

A first estimate of the effects that are to be expected can be obtained from the 
temporal evolution of the populations, $p(\e,t_w)$. 
In Fig.1a, I have plotted $p(\e,t_w)$ as a function of $\e$ for two final 
temperatures, $T=0.2\s$ (upper panel) and $T=0.4\s$ (lower panel), for various 
waiting times $t_w$. 
For $t_w\!=\!0$, one has a Gaussian centered at $\e\!=\!0$, cf. eq.(\ref{p.0}) and in 
the long-time limit, the center has moved to ${\bar\e}=-\b\s^2$.  
It is, however, obvious from the figure, that with increasing $t_w$ the distribution 
first narrows and then becomes broader again. 
Furthermore, the form of the distribution is not exactly Gaussian for intermediate 
times. 
A similar behavior of the distribution of inherent energies during aging has been 
found in a recent simulation of a binary Lennard-Jones system\cite{SVS04}. 
In a next step, I determined the various moments of the distributions, the mean 
value $\lg\e(t_w)\rg$, the width $\s(t_w)$ and the skewness $\g_1(t_w)$. 
The results are shown in Fig.1b. 
The change in sign of $\g_1(t_w)$ reflects the fact that the distributions are skewed 
to the high-energy side for small $t_w$ and to the low-energy side for long $t_w$. 
The deviations from a Gaussian are seen to be more pronounced for the lower final 
temperature. 
The fact that the distribution of $p(\e,t_w)$ becomes narrower for intermediate $t_w$ 
already gives rise to the expectation that the degree of non-exponentiality of the 
two-time CFs given in eq.(\ref{C.Q.L}) will change with the waiting time. 
This means that one expects the stretching parameter $\b_K$ of Kohlrausch fits 
($\propto\exp{[-(t/\t_K)^{\b_K}]}$) first to increase and then to decrease again.

In the following, the evolution of the intermediate scattering function and the 
rotational CFs after a quench from high temperature to several low temperatures will 
be discussed. 

\section*{III. Aging monitored by two-time correlation functions}
The equilibrium properties of the rotational CFs $g_L^{eq}(\t)$ for different 
values of $L$ have been discussed in detail in ref.\cite{elm98} along with the 
apparent translational enhancement.
Therefore, I will start with a brief discussion of the $Q$-dependence of the 
intermediate scattering function. 
In Fig.2a, $S_Q^{eq}(\t)$ is plotted versus $\t/\t_Q$ for various values of $Q$ 
and $T=0.3\s$. 
Here, $\t_Q$ is the relaxation time obtained from a fit to a stretched exponential 
function $\exp{(-(\t/\t_Q)^{\b_Q})}$. 
It is evident that only for small $Q$ one is in the hydrodynamic limit and the degree 
of non-exponentiality is a strong function of $Q$. 
In the free-energy landscape model, the reason for this behavior is explained by the 
fact that for small $Q$ many jumps have to be performed in order for the CF to decay. 
Therefore, $S_Q^{eq}(\t)$ averages over the density of states quite effectively and 
the decay is very slow and exponential. 
For large values of $Q$, any single jump decorrelates the intermediate scattering 
function and therefore there is no averaging. 
Consequently, the decay is governed by a broad distribution of effective relaxation 
rates. 
The situation thus is similar to the one for the rotational CFs for different values 
of $L$, c.f. the discussion on this point in ref.\cite{elm98}. 
For rotational CFs, however, the mean rotation angle enters separately, whereas due 
to $Q\!=\!q\!\cdot\!\d R$ the jump distance $\d R$ only multiplies the scattering 
vector. 

In Fig.2b the relaxation times $\t_Q$ and the stretching parameter $\b_Q$ from fits 
of $S_Q^{eq}(\t)$ to a stretched exponential function, $\exp{(-(\t/\t_Q)^{\b_Q})}$, 
are shown. 
For values of $Q$ smaller than roughly $10^{-2}$, the decay is exponential and one 
can define a diffusion coefficient $D_T\!=\!(\d\!R)^2\lg\k\rg/6$, where $\lg\k\rg$ is 
the average relaxation rate, $\lg\k\rg=\int\!d\e\k(\e)p^{eq}(\e)$. 
For larger $Q$, $\b_Q$ decreases and the relaxation time deviates from the 
$Q^{-2}$-scaling. For $Q>10$, neither $\t_Q$ nor $\b_Q$ change any further, because 
here one is in the limit of the 'jump correlation function'\cite{elm97}.

In Fig.2c, I have plotted the fitted relaxation times scaled to those at 
$T\!=\!0.9\s$ versus inverse temperature. 
The reason for the different behavior again lies in the different averaging over the 
transition rates for different $Q$. 
The maximum discrepancy between the relaxation times, however, does not exceed a 
factor of roughly two (three, if the average times are considered).
The values of $Q$ were chosen, because if $\d R$ is chosen between twenty and fifty 
percent of the van der Waals radius of a tetracene molecule, they are comparable to 
the scattering vector used in the mentioned experiment by Thurau and 
Ediger\cite{mark02}. 

Next, the CFs $S_Q(t_w+\t,t_w)$ and $g_L(t_w+\t,t_w)$ are calculated according to 
eq.(\ref{C.Q.L}), where now $t\!=\!t_w+\t$ is used as the time-variable. 
Results for different waiting times $t_w$ are shown in Fig.3a for $T=0.3\s$.
In case of the rotational CF, I used $L\!=\!1$, corresponding to dielectric 
relaxation. 
The results for other values of $L$ are quite similar, apart from minor quantitative 
differences that are relevant at equilibrium, cf. ref.\cite{elm98}. 
In all calculations, a mean rotation angle of $\Theta\!=\!10^\circ$ is used. 
This is on the order of magnitude as found experimentally for supercooled 
liquids\cite{BDHR01}. 
For long $t_w$, both CFs are independent of $t_w$ and coincide with those in 
equilibrium. 
It is evident, that the relaxation time as a function of $t_w$ changes much more for 
the rotational CF than for the intermediate scattering function. 
Also this finding is explained by the different inherent averaging over the 
density of states. 
As already pointed out above, for small $Q$, many $\e\!\to\!\e'$-transitions have to 
take place in order for $S_Q$ to decay. 
The same transitions are responsible for the approach of equilibrium. 
Therefore, the $t_w$ dependence of $\t_Q$ is less pronounced than the corresponding 
one for $\t_1$, the relaxation time decribing $g_1$. 
This is because in the latter case fewer $\e\!\to\!\e'$-transitions are required for 
$g_1$ to decay and the system exhibits less re-equilibration during the decay.

In Fig.3b, the results of Kohlrausch fits $\exp(-(t/\t_z(t_w))^{\b_z(t_w)})$, 
where $z$ denotes either $L\!=\!1$ or $Q$, to the CFs as shown in Fig.3a are 
collected. 
Two features are evident immediately.
The values of $\t_1(t_w)$ change by about $1.5$ decades from $t_w\!=\!0$ to 
$t_w\!=\!\infty$, wheras this change is much smaller for $\t_Q(t_w)$ ($0.7$ decades 
for $Q\!=\!0.1$ and $0.07$ decades fo $Q\!=\!0.01$). 
The stretching parameters show an increase for small $t_w$ and then decrease as 
equilibrium is reached.  
This 'hump' for intermediate $t_w$ has its origin in the narrowing of the 
distribution of the populations $p(\e,t_w)$, cf. Fig.1.
Thurau and Ediger\cite{mark02} found that $\t_{\rm rot}(t_w)$ changes by a larger 
amount as a function of $t_w$ than $\t_{\rm trans}(t_w)$ does.
In their interpretation of the experimental results, the value of the stretching 
parameter for the rotational CF was assumed to also show a slight increase followed 
by a decrease as a function of $t_w$.
Both effects are much more pronounced in the present model calculations, which might 
have its origin partly in the high initial temperature used here.
It should be mentioned, that the ratio $\t_1(t_w\!=\!\infty)/\t_1(t_w\!=\!0)$ 
strongly depends on the final temperature of the quench, for instance this ratio is
roughly $15$ and $6$ for $T\!=\!0.32\s$ and $T\!=\!0.35\s$, respectively.
As already mentioned, it will not be attempted to provide a quantitative description 
of existing experimental data.
The overall qualitative features are, however, very similar to what is observed 
experimentally.

Another important feature of the experiment, which supports the view of a spatially 
heterogeneous aging scenario is the fact that probe rotation needs longer to reach 
equilibrium than probe translation, at least for low final temperatures\cite{mark02}.
In Fig.3c, $\t_1(t_w)$ and $\t_Q(t_w)$ are plotted in a scaled way.
It is evident that the intermediate scattering function reaches equilibrium much 
faster than the rotational CF. 
In order to have a measure of the re-equilibration time, to be denoted as $\t_{eq}$, 
I have determined the $(1-1/e)$-points of the curves, cf. Fig.3c. 
These are plotted versus inverse temperature in Fig.3d, where I scaled them to their 
value at $T\!=\!0.5\s$. 
It is seen that the temperature-dependence of the re-equilibration times is somewhat 
different for translation and rotation. 
For lower temperatures, $\t_{eq}$(rot.) is longer than $\t_{eq}$(trans.) in 
qualitative agreement with the experimental results. 
The explanation of this finding again lies in the fact that the two CFs average in 
different ways over the density of states. 
Therefore, $S_Q(t_w+\t,t_w)$ reaches equilibrium faster than $g_1(t_w+\t,t_w)$.
It should be pointed out that in the framework of the present model it is the same
difference in the averaging over the density of states that gives rise to the 
apparent translational enhancement\cite{elm98}. 
Furthermore, the re-equilibration of course is determined by the $\a$-relaxation in 
the free-energy model as there is no other time scale.

\section*{IV. Aging in environmental fluctuation models} 
In the free-energy model the life-time of the dynamic heterogeneities is on the same 
time scale as the $\a$-relaxation time. 
This is because there is no other time scale in the model. 
In terms of Heuer's more general concept of a rate memory parameter $Q$ (not to be 
confused with a scattering vector) the free-energy model intrinsically corresponds to 
$Q\!=\!1$\cite{heuer97}. 
It has to be noted, however, that the free-energy model in the simple form 
presented here, cannot describe the finding of an exchange time much longer than the 
$\a$-relaxation time observed in a photobleaching experiment near $T_g$ in 
ortho-terphenyl\cite{CE95,mark.het}.
In this section, therefore, I will consider so-called environmental fluctuation 
models\cite{AU67, sill96} that allow for an extra time scale for exchange or 
arbitrary rate memory. 
In such models it is assumed that the molecular dynamics are faster in some part of 
the sample and slower in another. 
When applied to supercooled liquids the dynamic heterogeneities are related to a 
distribution of the correlation times and their life-time is limited by exchange 
among the regions of different mobility. 
Therefore, one would interpret the dynamic exchange as the structural relaxation of 
the system.  

Such models can also be used for the calculation of the aging properties of the 
relevant CFs. 
Here, I will concentrate on the rotational CF $g_1(t,t_w)$ and consider the following 
variant of an exchange model. 
It is assumed, that a reorientation rate $\G(\e)$ corresponds to each value of the 
variable $\e$. 
Therefore, the variable $\e$ in the present context characterizes a given 
environment and the value of the corresponding reorientation rate $\G(\e)$. 
These rates for simplicity are chosen according to 
\be\label{Gam.bp} 
\G(\e)=\G_\infty e^{\b\e} 
\ee 
The transition rates $\k(\e|\e')$ for $\e'\!\to\!\e$-transitions are again chosen 
according to eq.(\ref{kap.eps}). 
As mentioned above, it are these transition rates that are responsible for the 
exchange among the various reorientation rates in the sense that with every 
$\e'\!\to\!\e$-transition a change from $\G(\e')$ to $\G(\e)$ is accomponied. 
According to eq.(\ref{p.eq.eps}), one has a Gaussian distribution of $\e$-values and 
therefore a broad distribution of reorientation rates $\G(\e)$. 
This 'bare' distribution is narrowed somewhat due to exchange, if the latter takes 
place on a similar time scale as the reorientational dynamics. 
Note that the choice of $\G(\e)$ and $\k(\e|\e')$ are of a purely phenomenological 
nature. 
However, the details of the form of the distributions of reorientation rates and of 
exchange rates are of minor importance for the following qualitative discussion. 

The solution of the master equation for such exchange models is outlined in Appendix 
B where again it is assumed that reorientations proceed via rotational jumps with a 
mean jump angle $\Theta\!=\!10^\circ$. 
In the following, I will restrict myself to the assumption that each 
'exchange-transition' is accomponied by a randomization of the molecular orientation. 
In this case, the solution of eq.(\ref{Gl.BP}) is trivial (because of 
$c_L=\d_{L,0}$) and the corresponding rotational CF is given by $g_L(t,t_w)=\int\!d\e 
e^{-(\G_L(\e)+\k(\e))(t-t_w)}p(\e,t_w)$ with the reorientation rates 
$\G_L(\e)=\left[1-P_L(\cos{\Theta})\right]\G(\e)$, cf. eq.(\ref{G.L.eps}). 
The aging behavior of $g_1(t,t_w)$ is reflected solely in the populations $p(\e,t_w)$ 
which evolve in time according to the master equation (\ref{ME.eps}). 
These populations decay with effective reorientation rates $(\G_L(\e)+\k(\e))$, 
reflecting the fact that both, a reorientation as well as an 'exchange-transition' 
depletes the population $p(\e,t_w)$. 
In order to allow for different time scales of reorientations and exchange, I will 
use the following simple relation between the attempt frequencies: 
\be\label{k,inf.g.inf}
{\G'_\infty\over\k_\infty}=x
\quad\mbox{with}\quad\G'_\infty=\left[1-\cos{\Theta}\right]\G_\infty
\ee
Here, $x$ is a free parameter and the scaling to $(1-\cos{\Theta})$ assures that the 
reorientational geometry has no further impact on the relative time scales. 
For $x\!=\!1$, the time scale of molecular reorientation coincides with the time 
scale of dynamic exchange, whereas for larger $x$, the average exchange rates are 
smaller than the reorientation rates. 
Note that the assumption of a randomization of the molecular orientation means that 
also exchange gives rise to a decay of the rotational CF, which has some influence 
on its form for $x\!\sim\!1$ but not for large $x$ (slow exchange).

The rotational CFs $g_1(t,t_w)$ for $x\!=\!1$ behave very similar to those obtained 
in the free-energy landscape model. 
For large $x$, however, the re-equilibration takes very long.
This is a consequence of the fact that the aging behavior is determined by the time 
scale of exchange in this model. 
The $\e\!\to\!\e'$-transitions are responsible for both, the re-equilibration of the 
system and the exchange among the reorientation rates. 
For large $x$, the re-equilibration is still determined by the exchange time but the 
time scale for reorientations is faster. 
This behavior is summarized in Fig.4, where I have plotted the rotational correlation 
times $\t_1$ as a function of the waiting time for $T\!=\!0.4\s$ and 
$\Theta\!=\!10^\circ$. 
The re-equilibration for $x\!=\!10^3$ is much longer than for $x\!=\!1$. 
This behavior scales exactly with $x$ and it shows that in this model the 
re-equilibration is determined by the dynamic exchange and not by the reorientations. 

These model calculations show that within such an approach the assumption of an 
exchange time much longer than the $\a$-relaxation, as obtained in the mentioned 
optical experiments\cite{CE95,mark.het} near $T_g$, cannot be reproduced by assuming 
that dynamic exchange is governed by a structural relaxation which also determines 
the aging behavior of the system. 
This is because the re-equilibration time scale has been found to be determined by 
the $\a$-relaxation\cite{lunki05}. 
Therefore, one would have to include still another time scale, an 'aging-time', in 
such exchange models. 
This 'aging-time', however, would have to coincide with the $\a$-relaxation time. 

As already noted above, also the free-energy model in the simple form presented here 
cannot describe both features. 
It is not clear at present, whether the reason for this lies in the oversimplified 
assumptions regarding the geometry of molecular reorientations and if a more 
realistic description using a distribution of jump-angles\cite{hinze98} would allow 
to obtain longer exchange times. 

\section*{V. Conclusions}
In the present paper I have examined the out-of equilibrium behavior of the two-time 
CFs for translational and rotational degrees of freedom within the framework of a 
free-energy model for relaxation in glassforming liquids. 
The system has been prepared in an off-equilibrium situation by performing a quench 
from a high temperature to a low 'working temperature'. 
Instead of trying to obtain quantitative agreement with existing experimental data 
the focus has been on a qualitative description of the physics inherent in the model. 
Therefore, all model calculations were performed using the simplest choices possible 
for the various parameters involved. 
For instance, the transition rates were chosen according to a simple-minded rule, cf. 
eq.(\ref{kap.eps}), and the density of states was chosen to be Gaussian with a 
temperature-independent width. 
Both choices would have to be improved in order to obtain quantitative agreement with 
experimental results. 
In addition, I solely considered the generic situation of a quench from an infinitely 
high temperature. 

When considering the behavior of the intermediate scattering function and of the 
rotational CFs after a quench from a high temperature to a rather low temperature it 
is found that the correlation times of both functions increase continuously as a 
function of the waiting time that has elapsed between the quench and the 
measurement. 
For long waiting times, both correlation times reach their equilibrium values at the 
final temperature. 
However, the absolute change in correlation time is much larger in case of rotational 
motion than for translations monitored with small scattering vectors. 
In addition, the time needed to re-equilibrate, i.e. to reach equilibrium at the 
final temperature, also is longer for rotations than for translations. 
When considered as a function of the final temperature, the re-equilibration times 
$\t_{\rm eq}$ behave differently. 
For rotations, $\t_{\rm eq}$ shows a stronger temperature-dependences than for 
translational motions. 
These findings along with a waiting-time dependent stretching parameter are in 
qualitative agreement with experimental results by Thurau and Ediger\cite{mark02}. 

In the present model, the re-equilibration is determined solely by the 
$\a$-relaxation, as there exists no other time scale. 
This feature is in accord with recent dielectric experiments on a variety of 
supercooled liquids\cite{lunki05}. 
To allow for two distinct time scales I considered a so-called exchange model in 
addition to the free-energy landscape model. 
Such a model naturally provides two distinct time scales, one for on-site relaxation 
and one for dynamical exchange. 
It is found in the model calculations that $\t_{\rm eq}$ coincides with the 
equilibrium 'exchange time' if the exchange is associated with the structural 
relaxation responsible for the aging properties. 
Therefore, if the exchange time is much longer than the correlation time, 
$\t_{\rm eq}$ is much longer, too.
This means, in order to reproduce $\t_{\rm eq}$ on the order of $\t_{\a}$ in such 
models, yet another time scale has to be introduced. 

In summary, the free-energy model that has been introduced in order to describe the 
primary relaxation of supercooled liquids in equilibrium\cite{elm97,elm98} can also 
reproduce the experimentally observed features in out-of-equlibrium situations, at 
least on a qualitative level. 
The comparison with the exchange models shows, that aging experiments might also 
yield additional information about the nature of the dynamic heterogeneities in 
supercooled liquids. 
In particular, it appears an open question what the exact relation between the 
life-time of these heterogeneities and the re-equilibration time after a quench is. 
In the free-energy model the differences in the time scales, the correlation times, 
the life-time of the dynamic heterogeneities and $\t_{\rm eq}$ can unequivocally 
be attributed to the different averaging over the density of states for different 
dynamical variables. 
The underlying intrinsic time scale is determined solely by the $\a$-relaxation.
\vspace{0.2cm}
\subsection*{Acknowledgements:} 
I wish to thank R. B\"ohmer, B. Geil, G. Hinze and H. Sillescu for stimulating 
discussions.
Part of this work was supported by the DFG under Contract No. Di693/1-2.

\newpage
\begin{appendix}
\section*{Appendix A: Solution of the master equation for the free-energy landscape 
model}
\setcounter{equation}{0}
\renewcommand{\theequation}{A.\arabic{equation}}
In this Appendix I will give the relevant formulae that are needed for the
solution of the master equation for the composite Markov process $\{\e(t),a(t)\}$, 
where $a(t)$ has to be identified with the orientation $\Om$ or the position 
${\bf r}$. 
Because the solution of the master equation proceeds in the same way for both cases, 
thetreatment can be formulated quite generally. 
The transition rates can be written in the form 
\be\label{W.eps.a}
W(\e',a'|\e,a)=\k(\e'|\e)\L(a'|a)
\ee
Here, I neglected possible dependencies of $\L(a'|a)$ on the initial and final 
states of the transition. 
In the text, only the special case $\L(a'|a)\!=\!\d(a'-(a+\d a))$ is considered 
explicitly. 
To solve the corresponding master equation for this case one defines the matrix of 
the eigenvectors of $\L(a|a')$, $U(a,p)$ corresponding to the eigenvalue $\L(p)$, 
i.e. $\L(p)\!=\!\int\!da\int\!da'U^{-1}(a',p)\L(a'|a)U(a,p)$.  
Next, the conditional probability is expanded in terms of these eigenvectors, 
\be\label{G.eps.a}
G(\e,a,t|e_0,a_0)=
\int\!dpU(a,p)G_p(\e,t|\e_0)U^{-1}(a_0,p)
\ee
The Green's functions $G_p(\e,t|\e_0)$ are then found from the solution of 
\be\label{Gp.ME}
{\dot G_p}(\e,t|\e_0)= - \k(\e)G_p(\e,t|\e_0)
			  +\L(p)\int\!d\e'\k(\e|\e')G_p(\e',t|\e_0)
\ee
where the sum rule $\k(\e)\!:=\!\int\!d\e'\k(\e'|\e)$ was used for the diagonal 
element. 

The initial populations are chosen as described in the text, i.e.: 
\be\label{p0.eps.a}
p(\e,a,t\!=\!0)=p^{eq}(a)p(\e,t\!=\!0)
\ee
If one considers isotropic rotational motions with a fixed mean jump angle 
$\d\Om\!=\!\Theta$, one has to identify $a$ with $\Om$. 
In this case one has $p^{eq}(a)\!=\!{1\over 8\pi^2}$, 
$U(a,p)\!=\!\sqrt{{2L+1\over 8\pi^2}}D^{(L)}_{mn}(\Om)$, 
$\L(p)\!=\!P_L(\cos{(\Theta)})$ and the integration over $p$ is now a sum over $L$. 
In case of translational jumps onto all postions of a sphere with a radius $\d\!R$ 
(jump length), the corresponding substitutions in the general formulae are 
$p^{eq}(a)=V^{-1}$ with $V$ denoting the volume. 
Furthermore, one has $U(a,p)\!=\!{1\over\sqrt{V}}e^{i{\bf q}{\bf r}}$ 
and $\L(p)=j_0(q\d\!R)$, where $j_0(x)$ denotes the Bessel function of zeroth order. 

In eq.(\ref{Gp.ME}) one has to identify $p\!=\!L$ for rotations and 
$p\!=\!Q$ with $Q\!=\!(q\!\cdot\!\d\!R)$ for translations and thus eq.(\ref{Gp.ME}) 
explicitly reads: 
\Be\label{Gl.Gq.ME}
&&\hspace{-1.6cm}{\rm Rotation:}\hspace{1.0cm}
\dot{G}_L(\e,t|\e_0)=-\k(\e)G_L(\e,t|\e_0)
		       +P_L(\cos{(\Theta)})\int\!d\e'\k(\e|\e')G_L(\e',t|\e_0)
		       \nonumber\\
&&\hspace{-1.6cm}{\rm Translation:}\hspace{0.5cm}
\dot{G}_Q(\e,t|\e_0)=-\k(\e)G_Q(\e,t|\e_0)
		       +j_0(Q)\int\!d\e'\k(\e|\e')G_Q(\e',t|\e_0)
\Ee
Note that due to $P_0(x)\!=j_0(0)\!=\!1$, there is one eigenvalue, $p\!=\!0$, 
$\L(0)\!=\!1$.  
Therefore, for $p\!=\!0$, from eqns.(\ref{Gp.ME}, \ref{Gl.Gq.ME}) the original master 
equation for the transitions in the free-energy landscape, eq.(\ref{ME.eps}), is 
obtained, i.e. $G_0(\e,t|\e_0)\!\equiv\!G(\e,t|\e_0)$. 
This is in full accord with the idea underlying the model that the variable $a$ is 
allowed to change {\it only} in case of a $\e\!\to\!\e'$ transition\cite{elm98}. 
Once the $G_p(\e,t|\e_0)$ are obtained from the (numerical) solution of the 
equations (\ref{Gl.Gq.ME}), all quantities of interest can be calculated and one 
finds for the normalized correlation function: 
\be\label{C.p}
C_p(t,t_w)=\int\!d\e\!\int\!d\e'G_p(\e,t-t_w|\e')p(\e',t_w)
\ee
where $p(\e,t_w)=\int\!d\e'G(\e,t|\e')p(\e',t\!=\!0)$, i.e. only $p\!=\!0$ is 
relevant here. 
This is easy to understand from the fact that according to the chosen initial 
conditions, eq.(\ref{p0.eps.a}), the variable $a$ was in equilibrium in the 
beginning, $t_w\!=\!0$. 
In case of reorientational motions the normalization eliminates a factor 
$(2L+1)^{-1}$. 

In general, the behavior of the correlation functions is obtained from a numerical 
solution of the equations for the Green's functions, 
eq.(\ref{Gl.Gq.ME})\cite{elm97,elm98}. 
For this purpose, the transition rates $\k(\e'|\e)$ have to be chosen in a 
prescribed way, e.g. as done in the text, cf. eq.(\ref{kap.eps}). 
It is, however, possible to give some general results valid in specific situations. 
Consider the case of rotational random jumps, for which one has $\L(p\!>\!0)\!=\!0$.  
Similarly, if the incoherent scattering function is observed for large wave-vectors, 
$Q\!\gg\!1$ (or $q\!\gg\!(\d\!R)^{-1})$, one has $j_0(Q)\!\simeq\!0$. 

In the limit $\L(p)\!\to\!0$, the general expression for the Green's function,
eq.(\ref{Gp.ME}), approximately reads as 
${\dot G_p}(\e,t|\e_0)= - \k(\e)G_p(\e,t|\e_0)$ and therefore one has 
$G_p(\e,t|\e_0)\simeq\d(\e-\e_0)e^{-\k(\e)t}$. 
Using this expression, one easily finds from eq.(\ref{C.p}): 
\be\label{C.p0}
C_p(t,t_w)\simeq\int\!d\e e^{-\k(\e)(t-t_w)}p(\e,t_w)
\ee
which is independent of $p$ and gives the probability that the system has not left 
the metastable state with free-energy $\e$ occupied at $t\!=\!t_w$ in the time 
interval $(t-t_w)$. 

In the other extreme, namely very small wave-vectors or rotational diffusion,
one has $\L(p)\!=\!1-\eta_p$ with $\eta_Q\!=\!Q^2/6$ and 
$\eta_L\!=\!L(L+1)(\Theta/2)^2$. 
As has been shown in ref.\cite{elm98}, in equilibrium in the limit of small 
wave-vectors the model predicts an exponentially decaying intermediate incoherent 
scattering function, $S_Q^{eq}(\t)\!\simeq\!\exp{(-q^2D_T\t)}$, with an apparent 
diffusion coefficient $D_T\!=\!(\d\!R)^2\lg\k\rg/6$. 
In case of the rotational diffusion of molecules one accordingly has 
$g_L^{eq}(\t)\!\simeq\!\exp{(-L(L+1)D_R\t)}$ with $D_R\!=\!(\Theta/2)^2\lg\k\rg$. 
\section*{Appendix B: Environmental fluctuation model}
\setcounter{equation}{0}
\renewcommand{\theequation}{B.\arabic{equation}}
Such models can be defined by transition rates of the type\cite{sill96}
\be\label{W.BP}
W(\e',\Om'|\e,\Om)=W_\e(\Om'|\Om)\d(\e'-\e)+\k(\e'|\e)\L(\Om'|\Om)
\ee
where now the on-site reorientations are modelled via finite $W_\e(\Om|\Om')$.
Translational motions can be modeled in a similar way.
In the following, it is assumed that the reorientations proceed via rotational jumps 
with a mean jump angle $\Theta$ as in the free-energy landscape model. 
In addition, it has to be quantified what happens to the molecular orientation in 
case of $\e\!\to\!\e'$ transition. 
As in the original model of Beckert and Pfeifer\cite{BP65}, it will be assumed that 
either no change at all, $\L(\Om|\Om')\!=\!1$, or a random rotation, 
$\L(\Om|\Om')\!=\!1/(8\pi^2)$, takes place.
Proceeding exactly in the same way as in Appendix A, the Green's functions 
are obtained from:
\be\label{Gl.BP}
\dot{G}_L(\e,t|\e_0)=-\left[\G_L(\e)+\k(\e)\right]G_L(\e,t|\e_0)
		       +c_L\int\!d\e'\k(\e|\e')G_L(\e',t|\e_0)
\ee
where the reorientation rates $\G_L(\e)$ are given by
\be\label{G.L.eps}
\G_L(\e)=\left[1-P_L(\cos{\Theta})\right]\G(\e)
\ee
Additionally, I defined $c_L\!=\!1$, if $\L(\Om|\Om')\!=\!1$ and $c_L=\d_{L,0}$ if 
$\L(\Om|\Om')\!=\!1/(8\pi^2)$\cite{sill96}. 
As in the case of the energy landscape model, the relevant quantities are given by 
eqns.(\ref{C.p}) if again it is assumed that there is no correlation initially.
\end{appendix}

%
\section*{Figure captions}
\begin{description}
\item[Fig.1 : ] 
{\bf a}: The distribution of populations $p(\e,t_w)$ as a function of free-energy 
$\e$ for different values of the waiting time $t_w$ after a quench from
$T\!=\!\infty$ to $T\!=\!0.2\s$ (upper panel) 
($t_w\!=\!0,10^{-15},10^{-13},10^{-11},10^{-9},10^{-7},10^{-5},10^{-3},10^{-1},10^{2}
,10^4$) and $T\!=\!0.4\s$ (lower panel)
($t_w\!=\!0,10^{-5},10^{-3},10^{-1},10,10^4$).\\ 
{\bf b}: The mean value (upper panel), the width (middle panel) and the skewness 
(lower panel) of the distributions shown in (a). 
The skewness is defined as $\g_1(t_w)=\k_3(t_w)/\k_2(t_w)^{3/2}$ where $\k_2$ and 
$\k_3$ are the time-dependent cumulants.
\item[Fig.2 : ] 
{\bf a}: The intermediate scattering function in equilibrium ($t_w\!\to\!\infty$) for 
$T\!=\!0.3\s$ and different values of the reduced scattering vector 
$Q\!=\!q\!\cdot\!\d\!R$.\\ 
{\bf b}: Results of fits to a Kohlrausch function ($\exp{(-(t/\t_Q)^{\b_Q})}$) at 
$T\!=\!0.3\s$ as a function of $Q$. 
It is evident that $Q^2\t_Q\!=\!const.$ and $\b_Q\!=\!1$ holds only for small $Q$.\\  
{\bf c}: The parameters $\t_Q$ (upper panel) and $\b_Q$ (lower panel) as a function 
of temperature. 
The values of $\t_Q$ are scaled to their value at $T\!=\!0.9\s$, 
$\t_Q(T)_{\rm scaled}\!=\!\t_Q(T)/\t_Q(0.9)$. 
\item[Fig.3 : ] 
{\bf a}: $S_Q(t_w+\t,t_w)$ and $g_1(t_w+\t,t_w)$ versus 
$\t_{\rm scaled}\!=\!(\t/\t_Q)$ and $\t_{\rm scaled}\!=\!(\t/\t_1)$, respectively. 
Both panels show the correlation functions for reduced $t_w\!=\!10^{-7}, 10^{-1}, 1, 
10, 10^{6}\t_Q(\t_1)$. 
For the longest waiting time, the system has reached equilibrium.\\ 
{\bf b}: Results of Kohlrausch fits to $S_Q(t_w+\t,t_w)$ and $g_1(t_w+\t,t_w)$ as a 
function of the waiting time.\\ 
{\bf c}: The time scales $\t_Q(t_w)$ and $\t_1(t_w)$ versus $t_w$ are shown in a 
scaled way such that $\t_x(0)\!=\!0$ and $\t_x(\infty)\!=\!1$ for $x\!=\!Q,1$.\\ 
{\bf d}: The re-equilibration time, i.e. the time needed for the correlation time to 
reach its equilibrium value. 
The values of $\t_{\rm eq}$ are determined as the ($1-1/e$)-values of the curves 
shown in part (c).
\item[Fig.4 : ]
The rotational correlation time $\t_1(t_w)$ versus $(t_w/\t_1)$ for the exchange
model showing that for $x\!=\!10^3$ the re-equilibration takes much longer than the
rotational correlation time.
\end{description}
%
{\includegraphics[width=16cm]{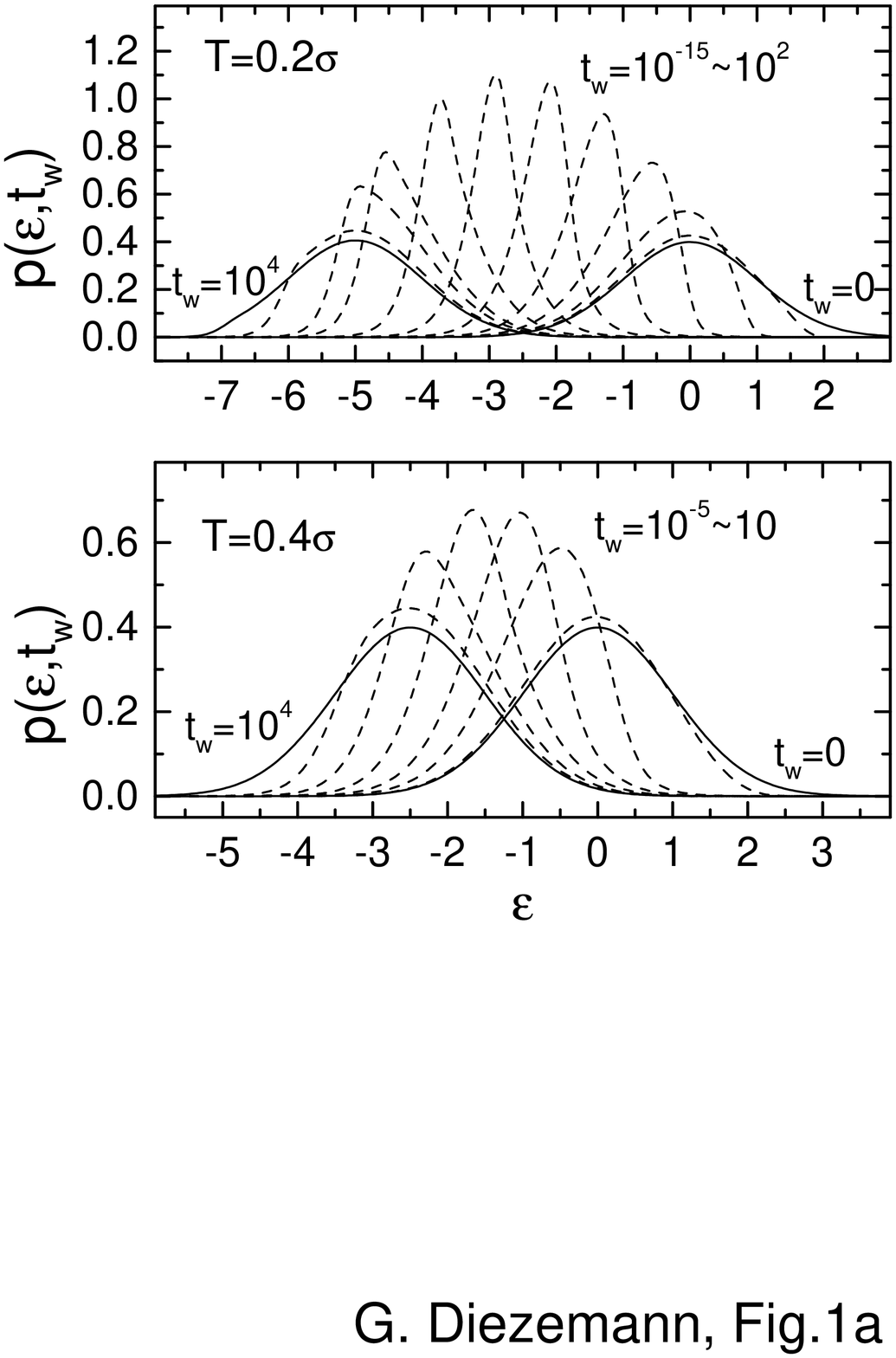}}
\newpage
{\includegraphics[width=16cm]{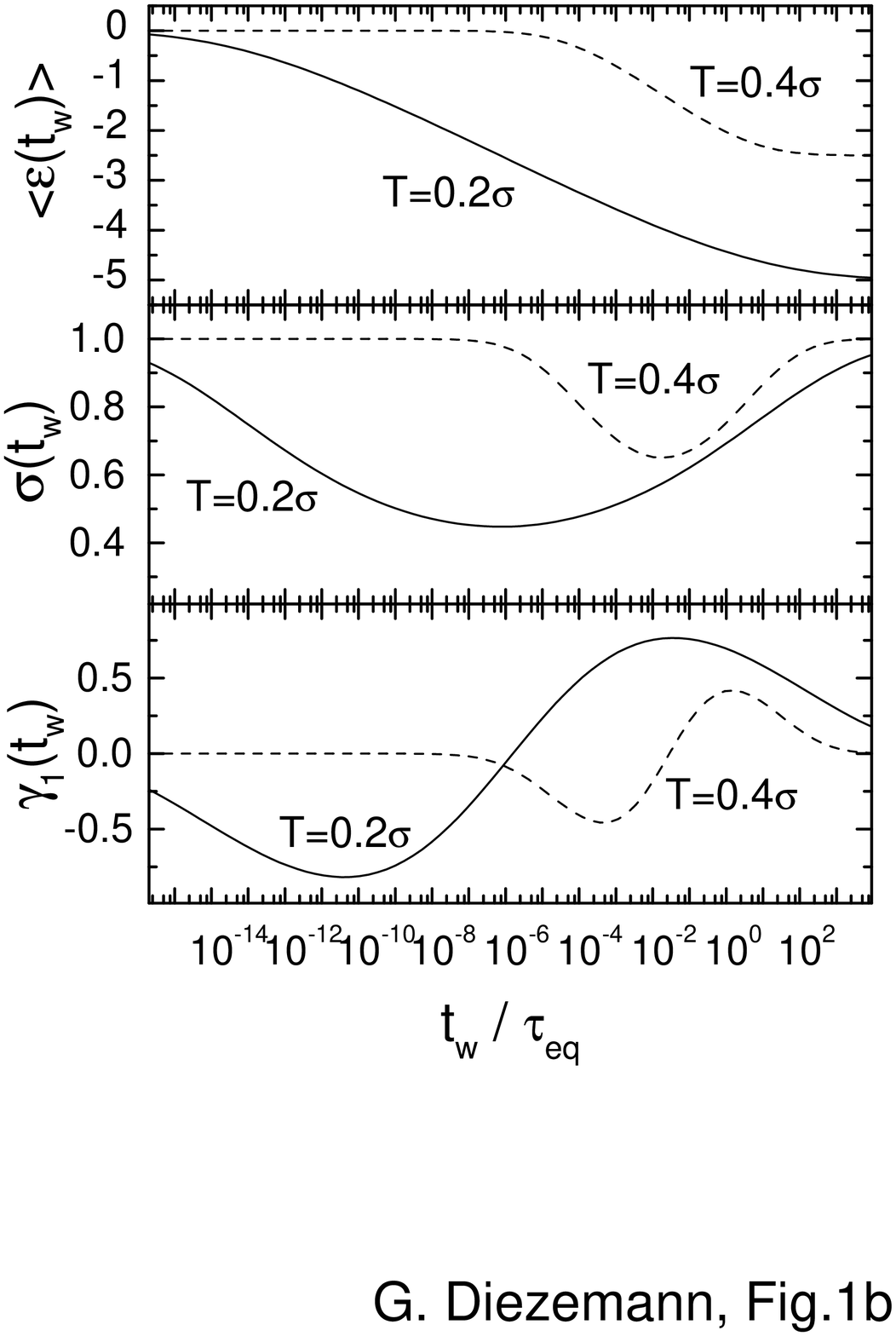}}
\newpage
{\includegraphics[width=16cm]{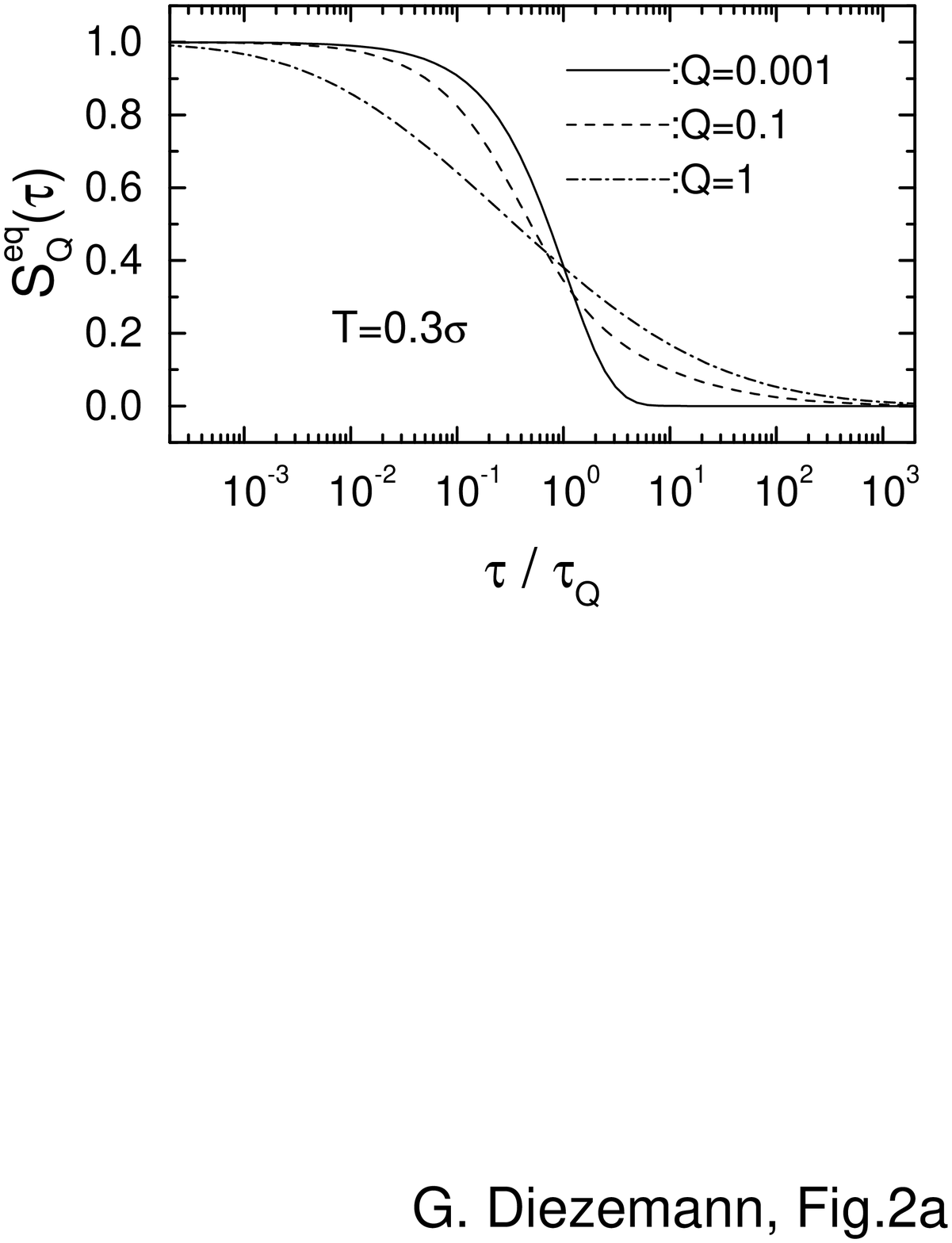}}
\newpage
{\includegraphics[width=16cm]{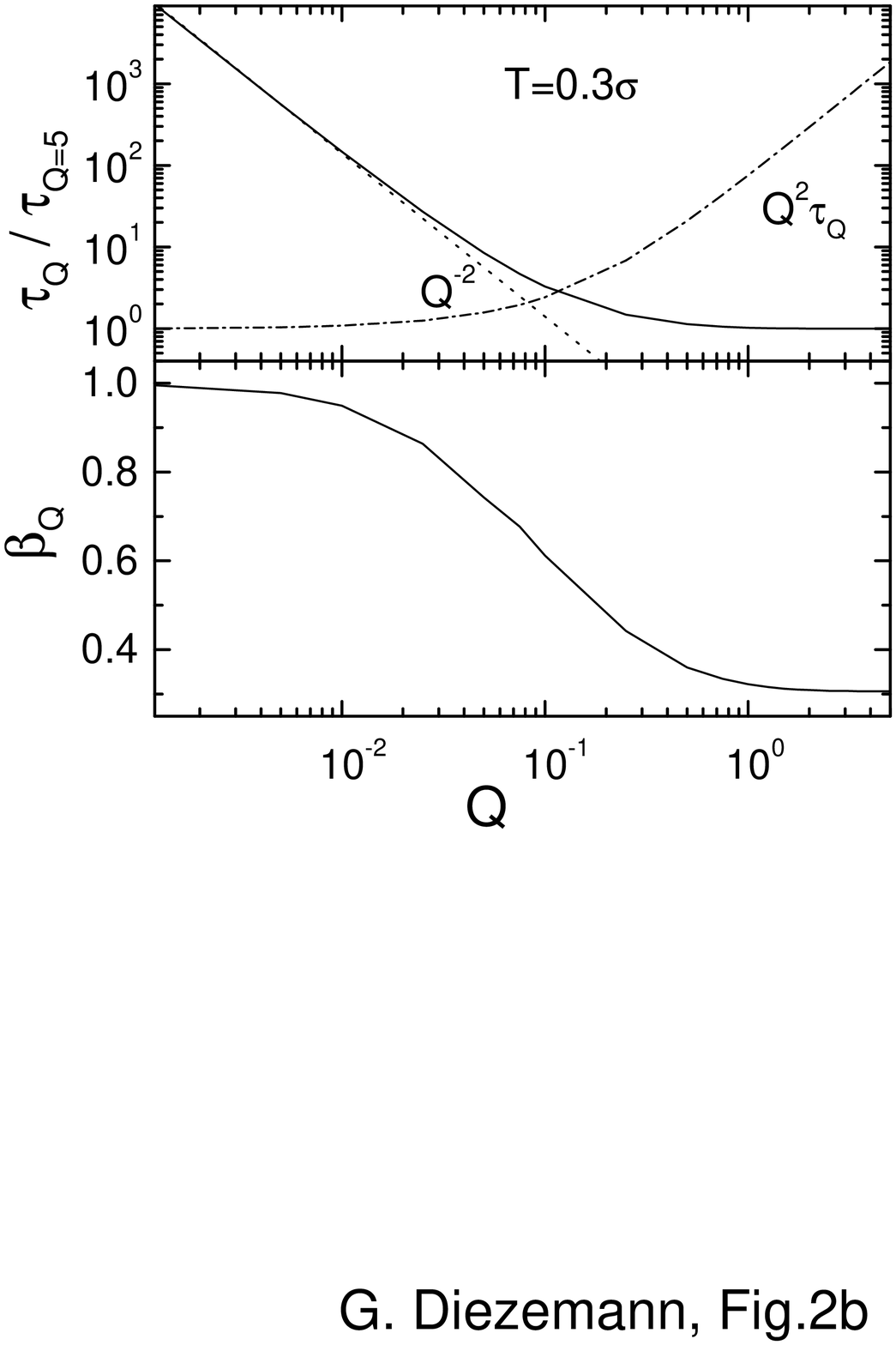}}
\newpage
{\includegraphics[width=16cm]{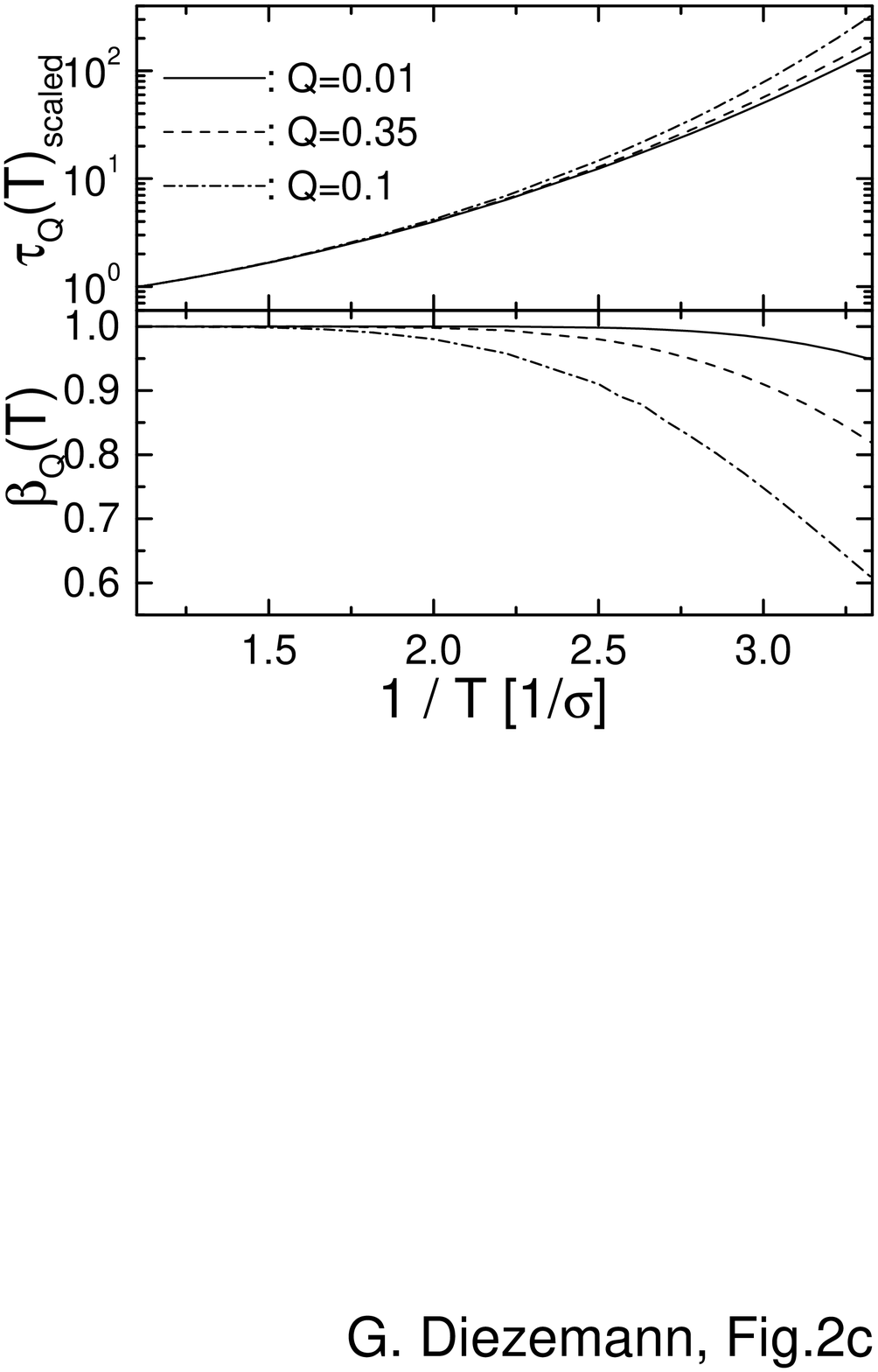}}
\newpage
{\includegraphics[width=16cm]{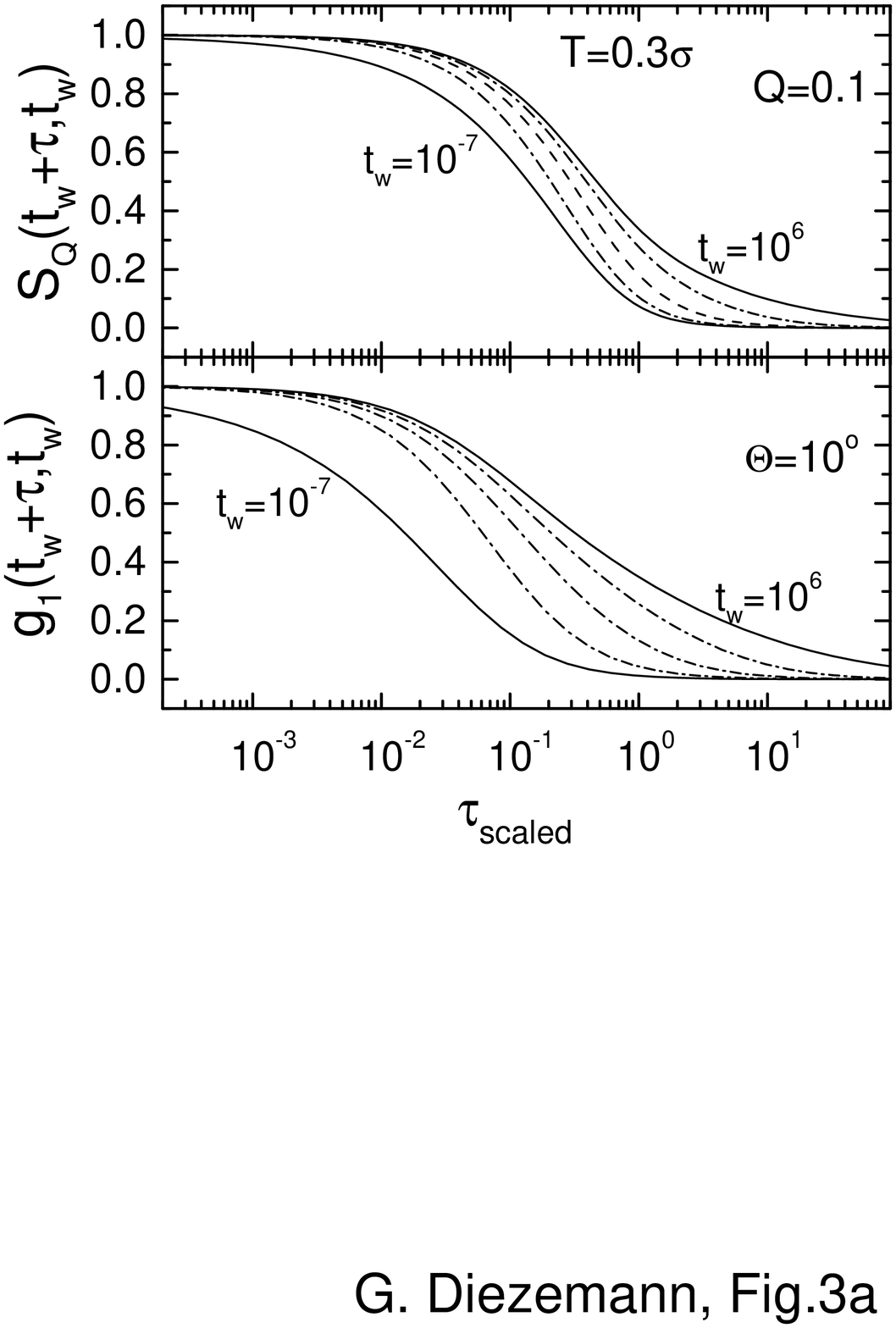}}
\newpage
{\includegraphics[width=16cm]{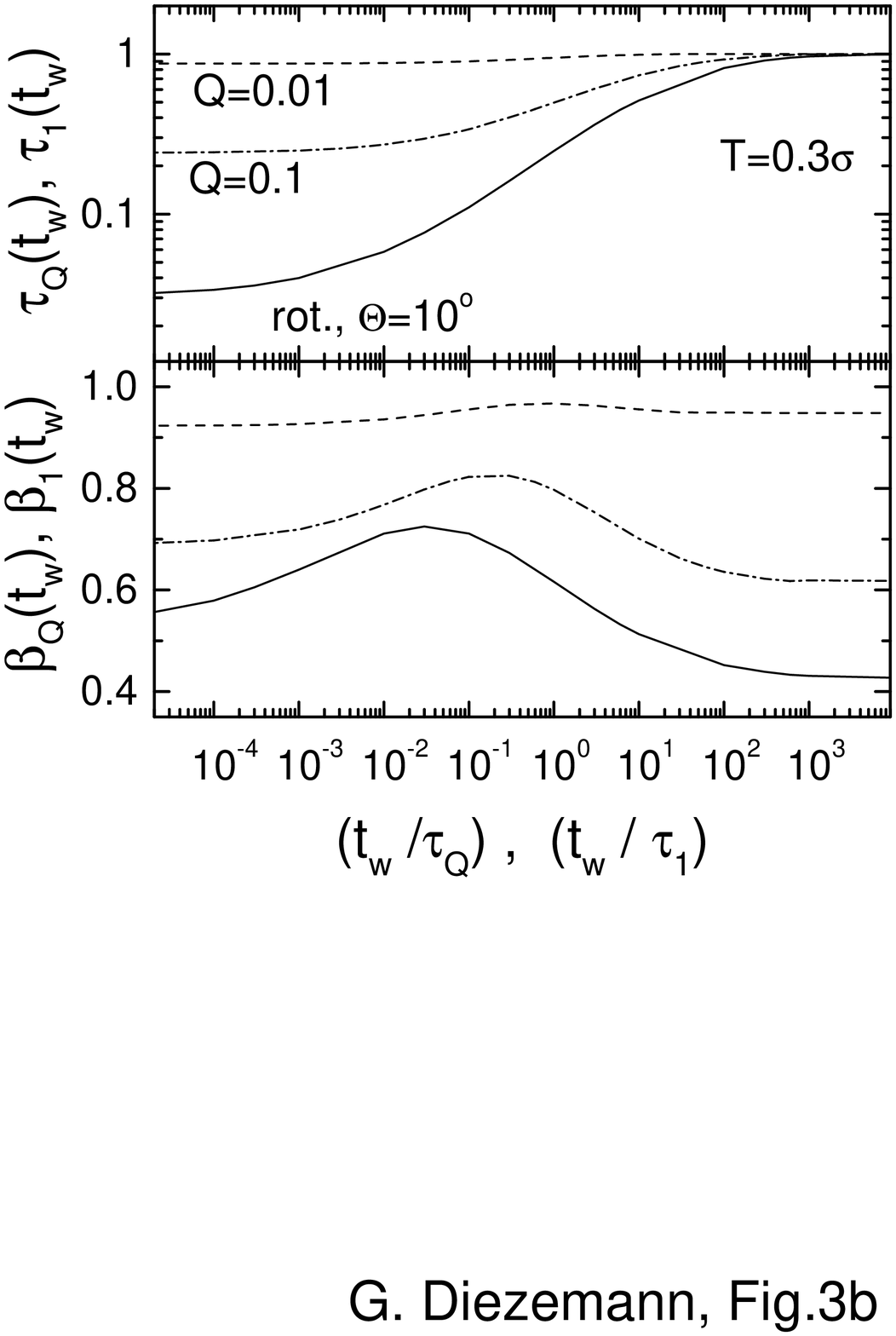}}
\newpage
{\includegraphics[width=16cm]{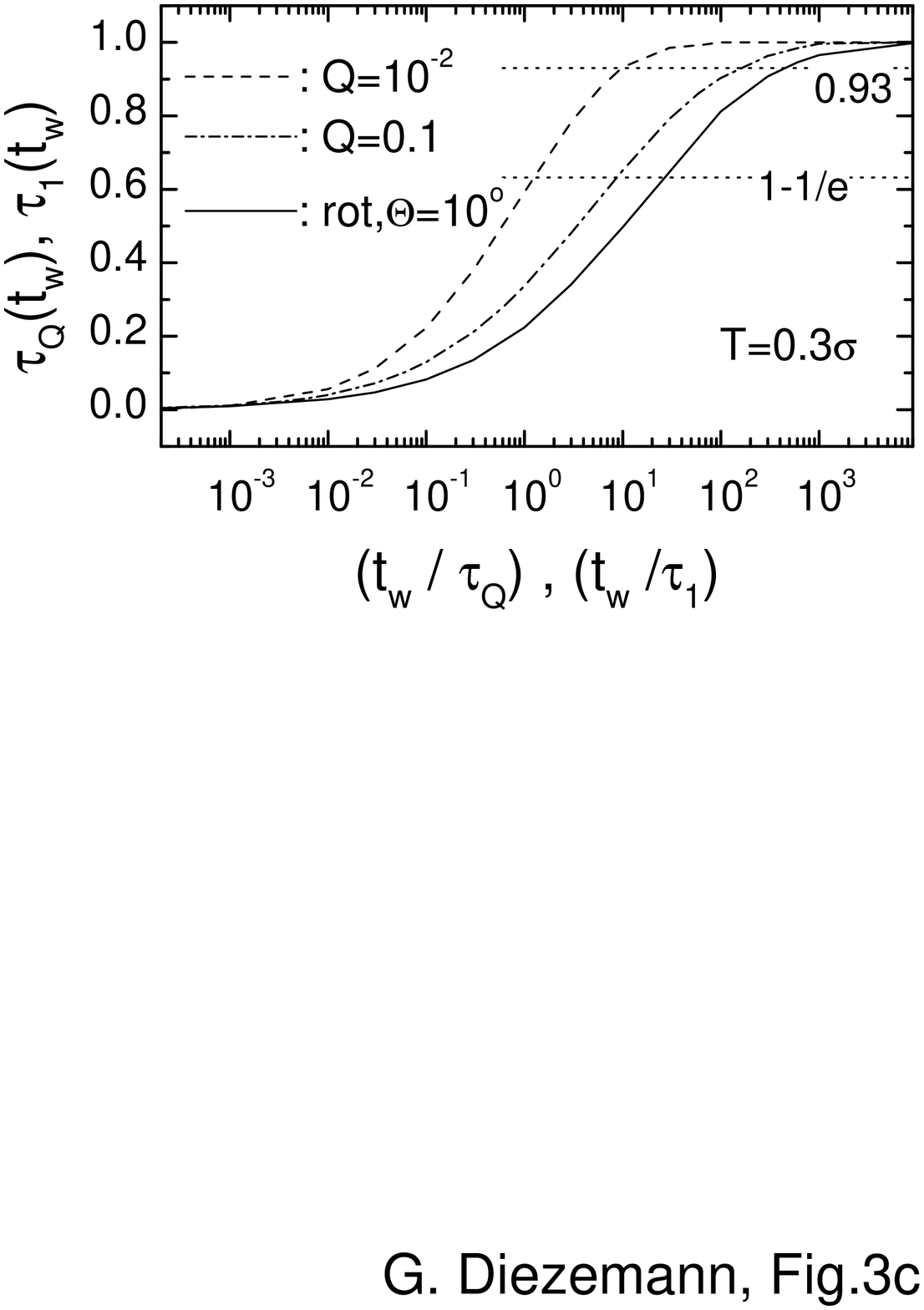}}
\newpage
{\includegraphics[width=16cm]{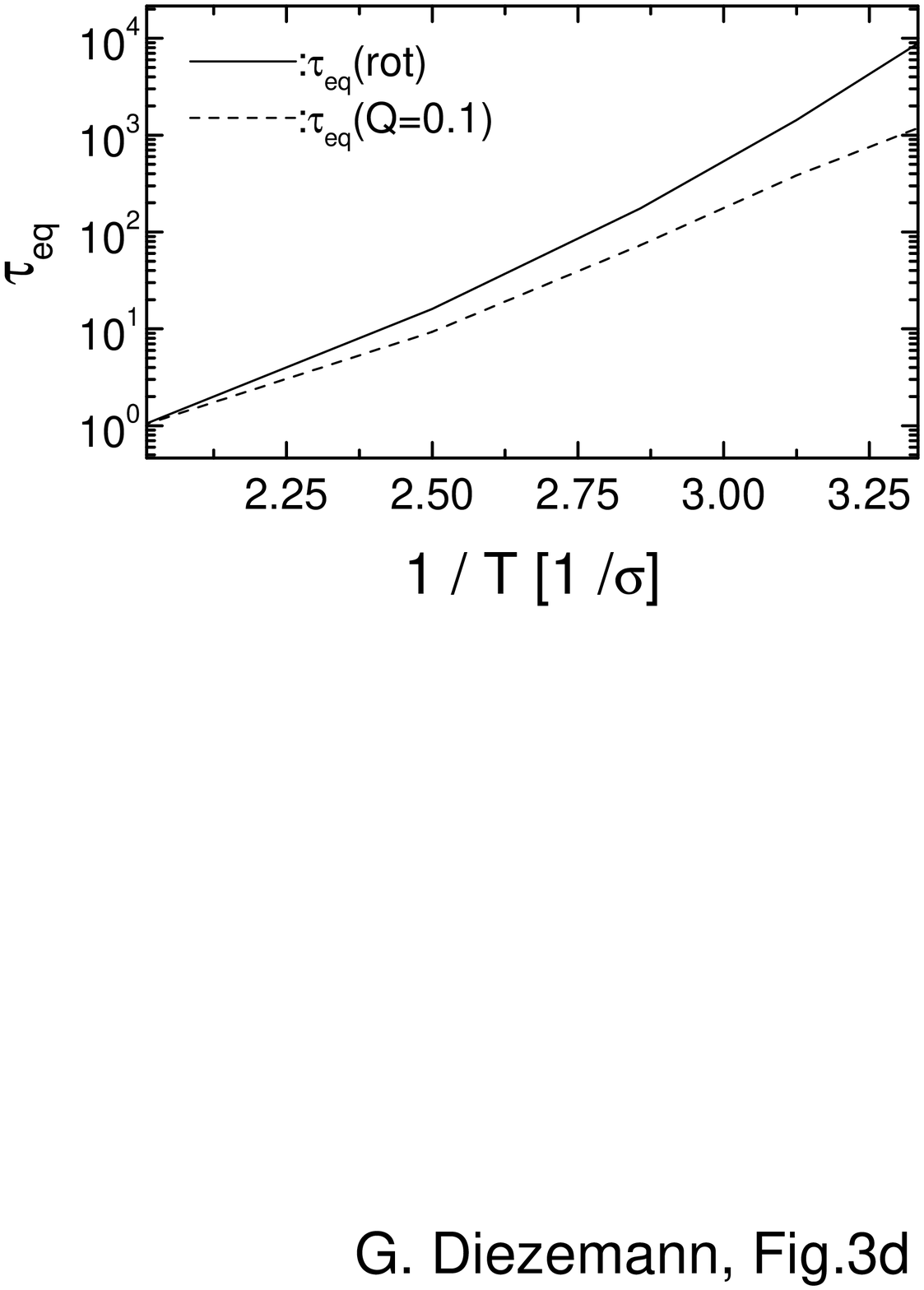}}
\newpage
{\includegraphics[width=16cm]{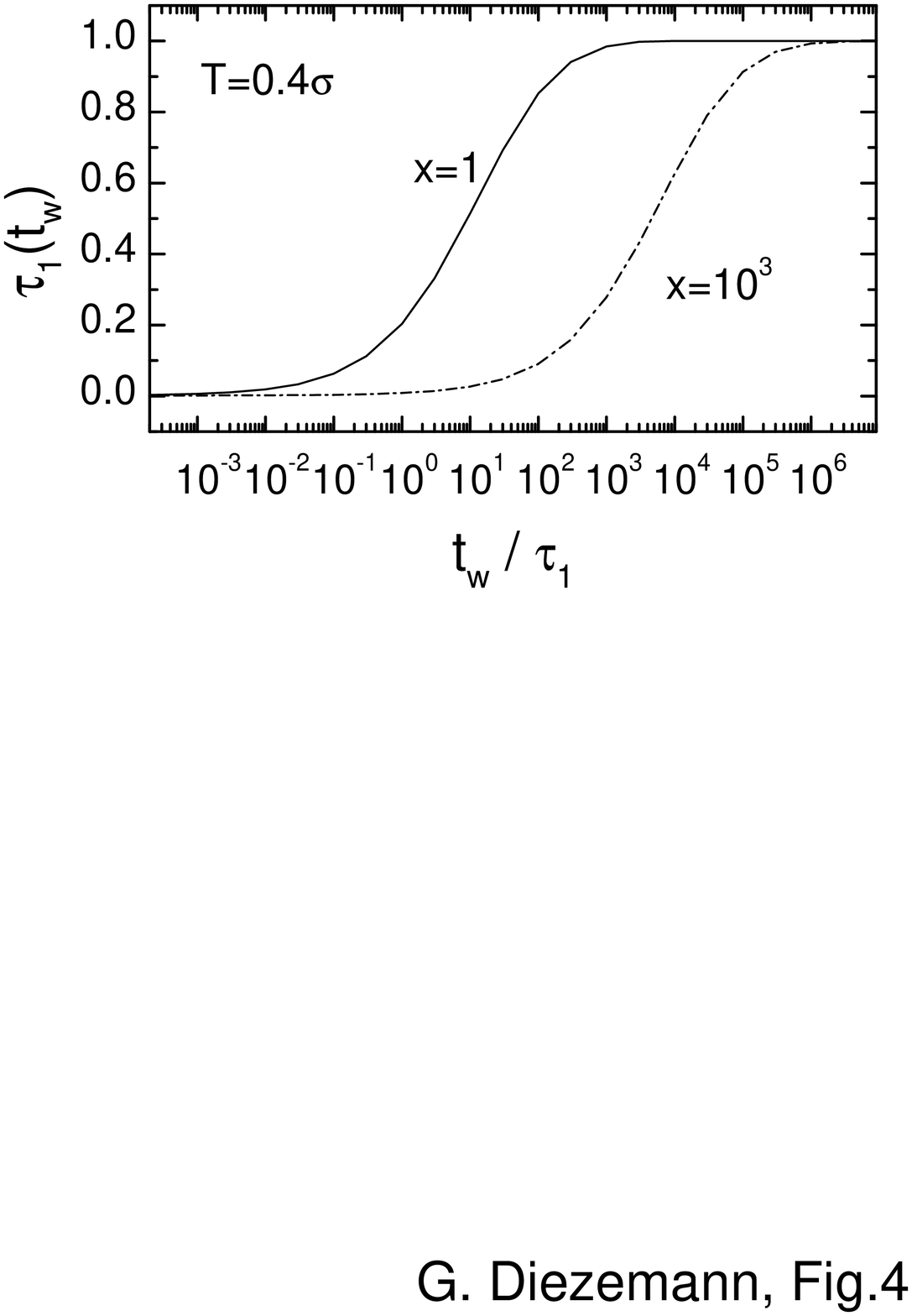}}
%
%
\end{document}